\documentclass[12pt,a4paper,english]{article}

\usepackage[brazil]{babel}
\usepackage{url}
\usepackage{amssymb,amsxtra} 
\usepackage{amsmath,color,graphicx}
\usepackage[utf8]{inputenc}

\usepackage{lineno}
\usepackage{indentfirst}
\RequirePackage{geometry}
\usepackage[T1]{fontenc}
\usepackage[bitstream-charter]{mathdesign}

\begin{document}

\begin{center}
  {\huge \bf In what sense space dimensionality can be used to cast light into cultural anthropology?}
\end{center}
\vspace*{0.25cm}

\begin{center}
  {\large\bf Francisco Caruso \& Roberto Moreira Xavier}
\end{center}
\begin{center}
  {\small \bf Centro Brasileiro de Pesquisas F\'{\i}sicas}
\end{center}

\begin{flushright}
\begin{minipage}{6.0cm}
\baselineskip=10pt
\vspace*{0.5cm}

\small{\textit{It is a pleasure to dedicate this work to our friend Francisco Antonio D\'{o}ria, scientist and philosopher, a man of multiple interests, in the hope that he will enjoy reading it.}}
\vspace*{0.5cm}

\end{minipage}
\end{flushright}

\section{Introduction}

Humans have always constructed spaces, through Mythos and Logos, as part of an aspiration to capture the essence of the changing world. This has been a permanent endeavour since the invention of language. By doing this, in fact, Humankind started constructing itself: we are beings in constant evolutionary process in real and imaginary spaces.  Our concepts of Space and our anthropological ideas, specially the fundamental concepts of \textit{subject} and \textit{subjectivity}, are intertwined and intimately connected. We believe that the great narratives about Humanity, which ultimately define our view of ourselves, are entangled with those concepts that Cassirer identified as the cornerstones of culture: \textit{space}, \textit{time}, and \textit{number}~\cite{Cassirer-Man}. To explore these ideas, the authors wrote an essay, in 2017, in a book format, in which the fundamental role of real and imaginary spaces (and especially of their dimensionalities) in the History of Culture was discussed. This book, titled \textit{O Livro, o Espa\c{c}o e a Natureza: Ensaio Sobre a Leitura do Mundo, as Muta\c{c}\~{o}es da Cultura e do Sujeito} \cite{nosso-livro}, has a preface written by Francisco Antonio D\'{o}ria. As many of the issues treated there are among his multiple interests, it was decided to revisit here the problems of subjectivity and subject's relationship with the dimensionality of space including the question of the architecture of books and other writing supports.

\section{A voyage through the History of Culture}

In Ref.~\cite{nosso-livro}, the story of a navigation through the intricacies of writing (and alphabets) -- in its different historical forms of registration --, by the jolts and rambles of Culture and by the extraordinary changes in Man's view of the world and of himself, is narrated. The first general idea was to highlight landmarks, traces, records of major changes in the human imaginary, in a view that can be associated with Fernand Braudel's ``longue dur\'{e}e''.

As an element of union and convergence between such apparently disparate narratives is the gradual construction of the concepts of space (real and imaginary). Special attention was given to space dimensionality and its influence on the perceptions and descriptions of the World and of Nature.

\textit{O Livro, o Espa\c{c}o e a Natureza} was conceived and built in order to shed some light on the confluence of the following questions:
\begin{itemize}
  \item Is there a direct relationship between the various manifestations of Culture and the physico-philosophical conceptions of Space?
  \item How does the different forms of writing registration affect or were affected by the current conceptions of Space?
  \item To what extent can the way of thinking be conditioned by the way of writing, thus influencing Culture and our understanding of Nature?
  \item How is human subjectivity and the notion of subject constructed based on the characteristic cultural soil of each era?
\end{itemize}

Ultimately, the authors hope to have contributed to consolidate the acts of reading and writing in a prominent place in the History of Ideas, by underlining their links with the construction of the absolutely fundamental concepts of space and time, as well as with the humankind subjectivity own formation.

The book \cite{nosso-livro} also aims to cast new light on philosophical anthropology, linking the issue of the subject to the dimensionality of real and imaginary spaces: from the verticality of the medieval theocentric being to the postmodern multidimensional being. In fact, so far the influence of space in anthropology is considered, we share Cassirer's position \textit{cum grano salis}. Indeed, we admit that one of the most important influences of \textit{space} concepts on Culture occurs through the dimensionality of the writing support: two-dimensionality, in the case of \textit{volumen}, and the three-dimensionality of \textit{codex}~\cite{nosso-livro}. On the other hand, the postmodern subject is characterized by the multidimensionality of digital networks, which tends to infinity. This particular original contribution will be the theme of our tribute to our dear friend Francisco Doria, on this very festive occasion.

\section{Subjectivity and Culture: highlights}

With the conquest of language, prehistoric man begins to deepen his place in the Universe, imagining and conceiving myths. Man and Nature are intertwined and gain meaning when thought as a whole. With the invention of Philosophy -- the appearance of \textit{Logos} -- \textit{Myths} are relegated to another plane and \textit{Physis} becomes the center of reflection on Nature, with early philosophers looking for the physical (material) principle of all things: the \textit{arkh\'{e}}. \textit{Logos} and \textit{Physis} are united and conceived together in the hands of the so-called physiologists. In the words of Karl Popper, ``Science must begin with myths, and with criticism of myths''~\cite{Popper}.

We cannot determine when the first man, in the face of unpredictable Nature and the starry sky, was taken by astonishment when realizing his own finiteness. Nevertheless, what we can affirm, with certainty, is that the perplexing problem of the origin and identity of Man appears in the oldest myths and has assumed a central place in Philosophy since its beginnings.

The issue of subjectivity and human identity is a task that demands a huge text which is completely out of our scope in this paper.
What we intend here is to shed some light on few aspects of the subjectivity formation, especially its relation to the history of the concepts of space and time.

Indeed, at decisive moments in his own history, Humans reorganize their view of the World~\cite{Arkan}, their relationship with the Universe, through a significant change in their space-time concepts. In those moments, the perception of themselves -- their identities -- also changes.

In order to simplify our narrative and make it intelligible in a few pages, we will make a brief summary of the landmarks of subjectivity history, from Classical Antiquity to the end of the \textit{Middle Ages}, then, to characterize the Modern Era, we will use the typology introduced by the sociologist Stuart Hall.

From the Greek philosophers Socrates and Plato, Man occupies the center of philosophical speculation.

Socrates is considered a landmark in the history of Greek Philosophy dividing it into two periods: the so-called pre-Socratic and the post-Socratic. Before him, physiologists were concerned with seeking the ultimate foundation of things, the primordial substances and the principles by which the plurality of \textit{Physis} results. With Socrates, Man becomes the main focus of Philosophy. In fact, Man is thought of as a social being, somehow inserted in a society that must be just and orderly (in the sense of the Greek word \textit{Kosmos}). For Socrates, \textit{Truth} is inside Man, but he cannot reach it alone. It is through \textit{dialogue}, understood as a way of exercising reason, that Truth will be discovered. Moral life is guided by thoughts which recognize the ideal values that are reached through Philosophy and Reason. Therefore, in Socratic philosophy, \textit{Man}, \textit{Ethics}, and \textit{Epist\'{e}me} are linked in a stable whole through the concept of \textit{Cosmos}.

In a very simplified way, for Socrates, \textit{Man is his "<Psych\'{e}">}. Happiness would result from the search of truth, which is sought through dialogue. Permanent questioning and inquiry are the way of building knowledge, always provisional. Thus, Man is a being who questions, dialogues, builds fleeting knowledge and concludes, with the Athenian philosopher: \textit{I only know that I know nothing}.

According to Plato, human nature is similar to a difficult text, the meaning of which must be deciphered by Philosophy~\cite{Apud}. Given that, for Plato and Socrates, Man is a being who dialogues in search of truth, thus he constantly needs the other, his mirror. The mirror metaphor is recurrent in Philosophy. Gaston Bachelard, for example, referring to the human capacity of dreaming, states that ``\textit{the dreamer cannot dream in front of a mirror that is not deep}''~\cite{Bachelard}, reinforcing the idea of a social being.

Aristotle starts from Plato's idea, but expands it, affirming, fundamentally, that \textit{man is his "<Nous">}. The term is commonly translated as \textit{intellect} (or also \textit{mind}, \textit{intelligence} and \textit{thought}). However, the translation, while correct in itself, leads the modern reader to a predominantly gnosiological and psychological problem, while the original semantic area in Greek includes a much broader problem: from ontology to metaphysics, from physics to cosmology, from anthropology to morality and even religion~\cite{Reale}.

Philosophical speculation about man continues its course in the Hellenistic period. With the rise of Christianity, Eastern influence is present. A great symbolism was built around the life and death of Jesus Christ, which has influenced the human attitude for a very long time, encompassing Late Antiquity and the entire Middle Ages. It should be emphasized that we refer to the fact that Christ presented his contemporaries with a new image of Man, much more based on internal attitudes than on external appearances, in addition to having especially valued \textit{love}, giving it at least the same weight as \textit{justice}~\cite{Ladner}.
\newpage

After a difficult period of consolidation, Christianity is adopted as the official religion of the Roman Empire by Constantine. At that time, the foundations on which the Western Cultural History will evolve are established: Greek Philosophy, Christianity and Roman Law. Saint Augustine is the philosopher who explicitly articulated this synthesis. It should be noted that this is the period in which the \textit{codex} is consolidated as a support for writing, as will be seen in Section~3. Thus, the conditions are created for the appearance of the \textit{medieval Man} (Section~4).

The crisis following the fall of the Roman Empire heralds a new era, a new mentality.
This moment -- the High Middle Ages -- sees the confluence of several ideas that are articulated with the radical experience of \textit{verticality}, in a religious society, turned to Heaven. That said, the three-dimensionality of space is associated with the concept of linear one-dimensional time, due to Saint Augustine, to form and reflect the emerging self-awareness of this \textit{theocentric man}. Indeed, the Middle Ages comprise a long period of time when religion defines all aspects of life -- family, work, the divine right to define social relations -- making society essentially theocentric: Men who live verticality build cathedrals. This verticality of being is contemporary with the three-dimensionality of the \textit{codex} as we shall see (Section~\ref{codex}). These interrelations, which emerge simultaneously, form the cultural soil from which a new and revolutionary mechanical vision of the world will be born, many centuries later~\cite{Dijk}.
That said, we can examine the emergence of the modern Man.

It is not difficult to realize that the multiple conceptions which Man has built about himself throughout History are so many and so varied that trying to synthesize them is practically impossible. However, there are some questions constantly revisited, since always: -- Is Man marked by his own personality or is he strongly influenced by life in common, by society? -- Is he essentially good, as Jean-Jacques Rousseau wants, or, as so much debated in the Christian world, does he bear the ``mark of Cain'' of capital sin, of wickedness?

Stuart Hall sees the history of Man in the modern period in an original way and tries to answer these questions. For him, modern Man, who announces the Renaissance~\cite{Renasc}, looks at the world and questions it -- in the manner of the ancients -- and is called by him \textit{enlightened subject} \cite{Hall}. He is significantly distinguished from the \textit{medieval man} for his openness, his new discoveries, marked by the valorization of what is \textit{new}: New World, New Science \cite{Vico}, New Art, in short, Man's new relationship with the Cosmos. This subject gains, later, new meanings with the ideals of freedom and equality of the French Revolution.

As social interactions intensify, Hall identifies the emergence of a new perception of man about himself and forges the expression \textit{sociological subject},who has consequently his identity formed from the interaction between individual and society.

This individual has an inner core which changes through dialogue that social life and the cultural world offer. This is the characteristic subject of the 19th century and the first half of the 20th century.

If the \textit{enlightened subject}, born in the Renaissance, according to Hall, saw the press invention and the generalization of the book, the Great Navigations, the Reformation, the Copernican and Newtonian Revolutions, the \textit{sociological subject} experiences the era of scientific, technological and social transformations of the 20th century. Today, we are witnessing his progressive disappearance and the birth of the \textit{postmodern subject} \cite{Harvey}, possessing a multifaceted identity that, in reality, can be considered as a superposition of several identities which, in some cases, can even be contradictory~\cite{nosso-livro}.
\newpage

Gradually, thinkers begin to realize that the influence of society is essential in the formation of Man. Perhaps the thinker who best expressed this view was Karl Marx. His conception was fragmentally exposed and later studied and systematized by his disciples, like Erich Fromm \cite{Fromm}. Supported by analysis of society after the Industrial Revolution, the Marxist conception of Man is based on two concepts: alienation and merchandise fetishism \cite{Marx}.

Indeed, the exacerbation of the characteristics of the \textit{sociological subject} by the impressive increase in social relations, resulting from the advent of the train, the telegraph, the radio and television, sets the stage for a new phase of Human history, that of the \textit{postmodern subject}. His characteristics appear little by little, but can only be clearly defined with the advent of internet, which makes men and women living beings in a multidimensional imaginary space.

In Section~\ref{post}, we will examine the impact of this radical change. First, however, let us examine the role of writing in different historical moments, encompassing these significant cultural revolutions, from Antiquity to the 20th century.

\section{From volume to codex}\label{codex}

Two major phases stand out in the history of Greek civilization. The first, in which \textit{orality} predominates, is Homer's world, and the second, marked by the \textit{affirmation of writing}, is the universe of Philosophy, based on Plato's work. This division is not a mere academic simplification, since the alphabetic transcriptions of Homer's texts can be seen as ``the beginning of a relationship between oral and written, a relationship that has proved fruitful''~\cite{Havelock}.

Such a relationship, in reality, leads us to the Platonic distinction between \textit{doxa} and \textit{epist\'{e}me}, which had an essential role in the dissemination of the written word, through the \textit{volumen}~\cite{Havelock-Plato}. Indeed, the pre-alphabetic oral and poetic tradition, characterized by \textit{parataxis} -- generically called by Havelock \textit{oral mental state} --, constituted the main obstacle to scientific rationalism, to the use of analysis, the classification of experience, its new systematization in the sequence of cause and effect. This is why the poetic mental disposition and, therefore, poetry, is for Plato the archenemy. It is this oral mental state that Plato associates with \textit{doxa}, considered to be an obstacle to \textit{epist\'{e}me} -- to scientific discourse -- associated with \textit{hypotaxis}, that is, to alphabetic culture.

However, it should be noted that the transformation of \textit{orality} into \textit{literality} in the classical period was a slow process. A beautiful work on this gradual change was made by Havelock~\cite{Havelock-Muse}. There, the author discusses the extent to which human consciousness itself changes, when culture becomes literate. He also discusses how this new form of communication affects the content itself and the meaning of the texts.

Plato and Aristotle's legacy will open new perspectives in the way of reflecting on \textit{Physis} and on Man, nested in a new \textit{Zeitgeist}. These philosophers are fundamental milestones in the History of Philosophy and in the reflection on Man. It was not at all an accident that, in the middle of the Renaissance, Raffaello Sanzio represented them at the center of his famous ``The School of Athens'', Plato taking the \textit{Timeo} in hand and Aristotle, his \textit{Ethics}.

From these two notable thinkers, Philosophy starts to be written. The way is open for the establishment of what can be called a new \textit{literary mental state}, which had significant consequences on human communication and directly reflecting on the function of \textit{volumen}. In practice, the philosophers under the influence of Plato's severe criticism of oral culture and the impact of Aristotle's great synthesis have contributed to the changes in the status of \textit{volumen}.

The \textit{volumen} -- and by extension, the library -- becomes, then, a welcoming space for different authors and, of course, for different arguments, in a new temporality; the diachronism of oral speeches is opposed to the synchronism of written texts~\cite{Jacob}.

In spite of this achievement, another support of writing was about to change Europe: the \textit{codex}.

Although there are records of some isolated appearances of the \textit{codex}, already in the first century AD ~\cite{Roberts,Cavallo}, it only spreads out from the third century onwards, and has a very particular meaning for Christianity, as material support of the Bible. According to Jacques Le Goff, ``Le Christianisme est une religion du Livre. Cette v\'{e}rit\'{e} n'a jamais \'{e}t\'{e} plus vraie que dans l'Occident m\'{e}di\'{e}val'' \cite{Bruc}.

The \textit{codex} is considered by some authors to be the first revolution in the history of book. Consisting of a sequential set of small groups of paper sheets sewn together, it has a format very similar to that of the modern book.

The material and practical advantages of \textit{codex} over the \textit{volumen} are many and significant. In general, its weight, much less than that of the parchment roll, makes it easier to handle. In addition, its shape also facilitates storage and it is no longer imperative to use both hands while reading, as required by the volume. You can also write on both sides of the sheet of paper. There is still a further but not insignificant advantage of facilitating random access to written work.

Some scholars attribute the adoption of \textit{codex} by Christians to purely economic reasons, which seems to us a simplistic way of minimizing the impact of the change in the mentality of that time. On this matter, we agree with \'{U}rsula Katzenstein \cite{Ursula} when she says that

\begin{quotation}
\baselineskip=11pt{{\small \noindent ``Perhaps (...) some (...) prominent person in ancient Christianity (...), no matter the ultimate purpose of his inspiration, managed to (...) imagine a different format for Christians manuscripts of the scriptures, which differentiated them both from the parchment scroll of Judaism and the papyrus scroll of the pagan world (...), imposing its use on all Christianity (...).
It was possibly an attempt by the ancient Christians to differentiate their writings from other literary forms, to mark them as sacred books.''}}
\end{quotation}

Over time, \textit{codex} takes on a very peculiar meaning, in addition to arousing superstitious respect among scholars during the Middle Ages. In short, the Book (the Bible) becomes the symbol \textit{par excellence} of the relationship between Man and God in a new Christian \textit{Weltanschauung}.

We have seen that Christianity's adoption of \textit{codex} articulates with the issue of \textit{verticality} or, in other words, with \textit{three-dimensionality of space}~\cite{nosso-livro}. It is through this third dimension that the \textit{medieval subject} is related to God. The \textit{codex} invention in the High Middle Ages occupies a prominent place in the formation of the emerging `` three-dimensional'' \textit{theocentric subject}. Therefore, a metaphor could synthesize this long historical period: \textit{the codex is the message}, to paraphrase McLuhan~\cite{McLuhan}. The essence of such an object which will be associated with Culture and its transmission for a long period is its three-dimensionality.

Here it is worth to emphasize that the awareness of the three-dimensionality of space and of the world, both real and imaginary, and the linearity of time pointed out by Saint Augustine are fundamental steps for the creation of Modern Science, many centuries later. These are the bases of a conceptual scenario in which movements and transformations can be studied and described by Modern Science. The greatest difficulty is to understand movement and its causes. Indeed, given the three-dimensional Euclidean space and linear time, the scenario of the World where Physics and History can be developed as possible sciences is set, but the understanding of motion will require, as we will show later, a new causal scheme, a complete revolution in Science and Culture, associated with the name of Isaac Newton, but also indebted to Copernicus and Galileo.

To conclude this Section, it is important to stress that this association between Space and God is not new. Actually, it is well known that Judeo-Christian conceptions of space were since early times identified with God. In fact, in 1st century Palestinian Judaism, the term \textit{makom} -- as the name of God -- was adopted for \textit{place}. A notable example of the confluence of stoic conceptions and Judeo-Christian beliefs around the nature of space and its identification with divine omnipresence was sustained by Iamblichus. The impact of this association is clearly seen, in a special way, in the evolution of mechanical theories in the Middle Ages, and, in the 17th and 18th centuries, culminating with the statement that space is nothing more than an attribute of God, or even identical to God. For example, for the English philosopher of the Platonic school of Cambridge, Henry More, space is the \textit{divine extension}, while, for Isaac Newton, absolute space is \textit{God's sensorium}.

\section{The medieval subject}\label{medieval}

From a philosophical point of view, Aristotle's Cosmology is consistent with the essence of Medieval Philosophy. Indeed, in Aristotelian Cosmos, there is a clear difference between the sublunar and the supralunar worlds. The highest world, the world of celestial bodies, is made of an incorruptible substance -- the \textit{ether}, or the \textit{quintessence}. In it, the movements of these bodies are eternal, because God moves the Universe. In our lowest world, underlining, everything is perishable, everything can decline and movements are no longer eternal. Those ideas are attractive since they are based on the judgment that the cause and origin of everything is the \textit{One}, the Absolute. From \textit{Unity} the variety takes place, in a type of degradation process.

The whole world is supported and hierarchically articulated by a kind of golden chain (\textit{aurea catena}). All material or spiritual things belong to this current. In this unifying thought there are two types of hierarchy: that of \textit{existence} and that of \textit{value}~\cite{Cassirer-Vesalius}. Since they are not opposed, there is a profound unity between medieval philosophy, morals and culture. Therefore, just as the two Aristotelian worlds are not made of the same substance and do not obey the same laws of motion, so the structures of the ecclesiastical, political and social world are subject to the same principle of hierarchy~\cite{Huizinga}. The feudal system is an image of this hierarchical system. It is necessary to understand how the hierarchical order of Aristotelian Cosmos -- in which the Earth occupies its center, where space is hierarchized and, in this hierarchical order, Man occupies the highest place -- is convenient for the Catholic Church and medieval society. The reason is that Christian doctrine, which permeated and dominated Middle Ages in the West, is based on the assumption of the existence of a general Providence governing the world and the destinations of Man.


The adoption of the \textit{codex} by Christianity marks the meeting of religious practice and meditation -- correlated with reading the Bible and Christian comments -- with the experience of Space. Perhaps the most striking example of this association is in new tendencies of Architecture towards the sky, with the construction of magnificent cathedrals~\cite{Lilley}. As for these, it is undeniable that they introduce verticality in the medieval cities in which they were built. In a dominantly ``shallow'' (plane) architecture, these imposing monuments stand out to reaffirm the importance and glory of God in those theocentric societies~\cite{Pano}.

In fact, it was Christianity that very soon adopted verticality, that is, it favored the upper-lower system, inspired by the resurrection and ascension of Christ, as well as the ascension of souls. Christian world and symbolic space thus become effectively three-dimensional. In the Middle Ages, as Jacques Le Goff teaches us~\cite{LeGoff},
\begin{quotation}
\baselineskip=11pt{{\small \noindent ``(...) this system will guide, through the spatialization of thought, the essential dialectics of Christian values. To ride up, to rise, to go higher, this is the stimulus of spiritual and moral life while the social norm is to continue in your place, where God placed you on earth, without having the ambition to escape your condition and taking care not to stoop, not to deprive oneself.''}}
\end{quotation}

In this way, a \textit{medieval subject} is shaped and trapped in his position in the theocentric society of that time, in a kind of effectively two-dimensional space, insofar he is prevented from escaping his social condition, leaving the third dimension as the only possibility of salvation.

These essential changes in the spatio-temporal categories of the Christian imaginary led, still according to Le Goff, to a slow but important process of belief in an intermediary kingdom between Paradise and Hell -- the
\textit{Purgatory} --, the third place. Expanding the geography of the beyond was an immense operation for Christians, because, in the last analysis, the inclusion of the Purgatory between the lower and upper kingdoms represented the adoption of the concept of an intermediary that corresponds, from the point of view of logical structures, to profound changes in social and mental reality in the Middle Ages.

It can be said, then, metaphorically, that the Christian world is three-dimensional not only because of the importance attributed to the \textit{upper-lower} axis in the medieval imaginary, but also because, in the Christian Middle Ages, binary logic is replaced by tertiary logic.

It is worth highlighting another general characteristic of the Late Middle Ages. The religious mental state slowly fades away and begins to give place to a new state of mind that will pave the way to the Renaissance and to the Scientific Revolution.

There is also an area of scientific knowledge, intrinsically related to the concept of Space -- and also to God -- which is about to change: it is the \textit{Medieval Cosmology}. The Medieval conceptions of Cosmos, dominated by the thought of Plato and Aristotle, will not resist Copernican criticism (Section~\ref{freud}). The Closed World has its days numbered.

A new subject emerges from the Renaissance, who came to be called by Stuart Hall \textit{enlightenment subject} -- as already said --, which, after all, results from an inextricable entanglement of concepts which are mirrored in the crucial question of the relationship of Man with the Cosmos has in the relations between subject and object. This implies two notable capacities of this new subject: that of coming to conceive the \textit{infinitude} of the Universe and of three-dimensional space and that of not succumbing to the amazement caused by this non-divine infinity (Section~\ref{freud}). The \textit{enlightened subject} does not become small. On the contrary, as has already been seen, he grows, from the moment when he realizes how much, through his intellect, he is perfectly capable of embracing, conceiving and rationalizing all these questions that touch infinity, from the microcosm to the macrocosm. Perhaps this is why Ren\'{e} Descartes places the soul, the intellect -- \textit{res cogitans} -- at the core of his view of the subject, as opposed to \textit{res extensa}.

\section{Freud's first narcissistic wound and the new subject}\label{freud}

The publication of \textit{De Revolutionibus Orbium Coelestium} by Copernicus was a hard stroke to \textit{medieval subject}, who believed himself made up in the image and likeness of God and a privileged occupant of the center of Universe. He suddenly becomes a peripheral being, as the Earth, following Copernicus, comes to be seen only as one of the planets that revolve around the Sun. This impact on the collective imagination became known as the first of the so-called three narcissistic wounds, an expression coined by Sigmund Freud. The other two wounds refer to the contributions by Charles Darwin and Freud himself.

Alexandre Koyr\'{e} goes so far as to claim that the publication of Copernicus' greatest work marks the end of the Middle Ages. In his words \cite{Koyre},
\begin{quotation}
\baselineskip=11pt{{\small \noindent ``The year 1543, the year of the publication of the \textit{De Revolutionibus Orbium Coelestium} and the death of the author, Copernicus, marks an important date in the history of human thought. We are tempted to consider this date to mean the end of the Middle Ages and the beginning of modern times, because, more than the conquest of Constantinople by the Turks or the discovery of America by Christopher Columbus, it symbolizes the end of a world and the beginning of other.''}}
\end{quotation}

Copernicus' heliocentric system calls into question the unity of Aristotle's Physics. The unification between the physical description of motion on Earth (or in the sublunar world) and in Heaven is shaken because Aristotelian Physics, Astronomy and Cosmology depend on the hypothesis that the Earth is the center of the Universe. In the Copernican system, Astronomy becomes heliocentric, while physics remains geocentric. This fact has an immense importance in the history of ideas at the Renaissance. For instance, the Italian philosopher and historian of science Paolo Rossi, referring to the Copernican system, says that~\cite{Rossi}

\begin{quotation}
\baselineskip=11pt{{\small \noindent ``the admission of the terrestrial movement and the acceptance of the new system involved not only a reversal of the astronomical structure and physics, but also a modification of current ideas about the world, a new assessment of the place and meaning of man in the Universe.''}}
\end{quotation}

There is no doubt that the Copernican Revolution scared a lot by revealing a new spatial scale. Although, in fact, the immediate result of the publication of \textit{De Revolutionibus} has been to spread skepticism and disturbance, the theory exposed in it proved to be more effective in the long run.

However, the Copernican Cosmos is still finite. Copernicus gives movement to the Earth and claims that the visible world -- the world of fixed stars -- is not measurable, it is \textit{immensum}. Koyr\'{e} comments that, in a way, it is curious that Copernicus gave the first step to think of an infinite World by stopping the movement of the material orb of the fixed stars, but has hesitated to give the second, that of dissolving this sphere in boundless space.

Anyway, shortly afterwards, some Copernicans took the step that the Polish astronomer did not take, declaring that the sphere of fixed stars does not exist. ``In the starry skies the stars were located at different distances from Earth and that these skies «stretched infinitely upwards».''

Inspired by Copernican ideas, Giordano Bruno stated categorically the infinity of the Universe, with the possibility of life in other planets~\cite{Duhem}.

About this possible infinite world, Blaise Pascal will say later, in the 17th century: ``Le silence \'{e}ternel de ces espaces infinis m’effraie''~\cite{Pascal}.

Returning to the historical perspective, we would now like to emphasize that, in our opinion, Andreas Vesalius' contribution to Anatomy goes to the opposite direction of that of Copernicus, regarding the position of Man in the Renaissance World, \textit{vis-\`{a}-vis} Science.

Vesalius, in the same way as Galileo, was engaged in an ongoing struggle against philosophical authorities. In the same year that \textit{De Revolutionibus} comes to light, his classic work \textit{De humani corporis fabrica} is published. He did not give space in his work for metaphysical speculations. In it, to the detriment of valuing abstract theories, there is a clear valuation of observations and experiments, made from the dissection of corpses, in the best Leonardian spirit. The Belgian doctor's work has an interest that transcends the particular; its merit, in addition to being scientific, is also philosophical. Indeed, Vesalius removes Anatomy from medieval mystical speculations, while, also as an act of rupture, Galileo will seek to leave Science away from theological disputes.

On the other hand, it should be remembered that, in the Middle Ages, Anatomy was considered a ``minor'' activity, as it dealt with the dead and required hard (and somewhat repulsive) practical work. Vesalius' strict observance in solving his problems in perfect conformity with experience will give Anatomy a new status, and lead him to follow this new path of intellectual honesty suggested by Leonardo da Vinci, bringing Anatomy and, consequently, Man and Human body, to the core of scientific debate in the Renaissance.

Therefore, one can say that the year 1543 witnesses two antagonistic movements: one in which Man is put aside by Copernicus (as a corollary of his Astronomical Theory) and another that brings Man's body to the center of  interest of Medicine.

It is worth noting that many authors consider that the Cartesian idea in which \textit{res cogitans} is opposed to \textit{res extensa} is an expression of these opposite movements and marks the birth of the Subject and the notion of subjectivity. We cannot forget that, if Descartes puts the intellect at the center of his reflection, Vesalius makes the human body the central object of the study of life. Thus, it is through the effort of these two thinkers that Man and Subjectivity come to occupy the center of the new Worldview.

The formation of the \textit{enlightened subject} depends also on a central problem of Renaissance, namely that of \textit{self-consciousness}. Thinkers, like the Italian poet and writer Francesco Petrarca, considered to be the father of Humanism, the German cardinal Nicholas of Cusa, the Italian Neoplatonic Giovanni Pico della Mirandola and the Italian philosopher, Marsilio Ficino, another exponent of Humanism, contribute to a new personality ideal of the Renaissance, to the formation of a \textit{new subject}. Man begins to relate to the world no longer as in \textit{Middle Ages} and this change is related to the general problem of Renaissance involving the relationship between \textit{subject} and \textit{object}. In Cassirer's opinion,
\begin{quotation}
\baselineskip=11pt{{\small \noindent ``Man is for the universe, self is for the world just as the contained is for the continent. Both determinations are equally essential to express the relationship between man and cosmos. Thus, there is a constant reciprocity between them, the constant transformation of one into another.''}}
\end{quotation}

According to Ficino, human action in all its manifestations -- artistic, technical, philosophical or religious -- basically expresses the divine presence of an infinite \textit{mens} (mind) in Nature. Man is exalted as a
\textit{microcosm}, synthesis of the Universe. This view is not new, as it was already discussed by Plato and the Neoplatonists, but it will acquire new dimensions in the Renaissance and, especially, in the Enlightenment.

A beautiful example of this relationship between God and Man, different from that found in the Middle Ages, is implicit in the artistic representation of the \textit{Vitruvian Man} by the brilliant Italian painter and scientist Leonardo da Vinci, made around 1490.

In this famous da Vinci's drawing, a naked man is seen represented simultaneously in two different positions with the arms inscribed in a circle and in a square. It is well known, since Classical Antiquity, that the \textit{circle} is considered a symbol of perfect movements, of divine perfection or, in short, of God himself. The other geometric figure used in the representation of this Vitruvian Man is the \textit{square} which, unlike \textit{circle}, is not a form that appears in nature; it is, therefore, a construction of the human mind.

By simultaneously using the two geometric figures to inscribe the same man in two different positions, Leonardo is addressing the old question of \textit{squaring the circle}, but not only from a mathematical or even a technical point of view. The Italian genius, in this drawing, is telling us two complementary things about Man: on the one hand, that the (golden) proportions of the human body reflect the divine character of the Creator; on the other, that this same Man may have something divine, since he is able to perceive in Nature the presence of a divine \textit{mens}. God offers the circle, and Man, the square, as concatenating forms of
\textit{order}. God permeates the entire Book of Nature, from which subjects read natural relations and apprehend Nature from Geometry. It is through the intellect and a geometric framework of the world, therefore, that this new Man can rise to God. In this way, microcosm and macrocosm would be related, in the same way that Man is in harmony with God. We can foresee here traces of Baruch Spinoza's maxim \textit{Deus sive Natura}.

\section{The postmodern subject}\label{post}

The substitution, or rather, the tendency to replace the paper world by the digital universe, seen through the screen attached to a computer, was the starting point for the reflection of several authors on the future of the dissemination of written texts; whether or not a revolution in the art of reading and writing will result, only History will tell us. If the computer effectively replaces the book, we will have a new order of spatial imagery, with reflexes, still unknown, on the forms of contemporary thought, as the French historian Roger Chartier warns us, for example~\cite{ordre}:
\begin{quotation}
\baselineskip=11pt{{\small \noindent ``the universe of electronic texts will necessarily mean a departure from mental representations and intellectual operations specifically linked to the forms that the book had in the West seventeen or eighteen centuries ago. No order of discourse is, in fact, separable from the order of books that is contemporary with it.''}}
\end{quotation}

Even though there are still no signs that the book will have the same fate as the \textit{volumen} and the \textit{codex} -- on the contrary, it is known that the computer increased the amount of written texts on paper --, some significant changes in culture have undeniably been introduced by the computer.

In the mazes of networks~\cite{redes}, in the digital universe of the \textit{internet}, a new type of random access is changing the relationship between Man and the text, favoring subjectivity, subjective interpretations and fragmentation of reading, which will bring unexpected consequences, as we will see. It also brings back the dream of a \textit{bibliotheca universalis}, now without walls or borders~\cite{Bloch}. The \textit{book} had solved the problem of \textit{temporality} of information; \textit{computer} will solve the problem of \textit{spatialization} of information. Indeed, the book allows the reader to ``dialogue'' with writers from different eras, but the consulted books must be in the reader's hands, whether at home, at work or in a library. Reader and printed book must be necessarily in the same space. With access to digital books stored in virtual libraries, through the computer (or modern cell phones) and the internet, the reader can be anywhere in the world.

New digital technologies also allow for a new treatment and diffusion of images, particularly notable in Medicine and Science, as well as in the fields of Art and Communication~\cite{Lister}.

In summary, video clips in the universe of the new media -- which fragments space-time -- may be considered the perfect symbol of a society in which order, memory and the causal connection of facts are undervalued. To this corresponds, in Science, the interest in complex systems -- chaos and fractal geometry -- and the abandonment of the Cartesian program, perhaps an omen of the formation of a new mentality, of a new \textit{postmodern mental state}.

The postmodern, contemporary subject synthesizes in his body and mind the very history of Man. He keeps the ancestral memory of all the perplexities that have built us and results, essentially, from the exacerbation of the social ties that mark men and women of the 20th century.

In reality, these social ties grew exponentially with the advent of radio and television. This fact has been technically and socially determining for the \textit{global village} utopia and for the revolution that follows with computers and internet.

One of the most remarkable legacies of the computer and the computer age was the fact that, in 1990, the British physicist Tim Berners-Lee developed at CERN, the \textit{World Wide Web}, or simply Web or even www, which is nothing more than a system of hypermedia documents that are interconnected and executed on the Internet, articulated in a \textit{cyberspace}. This network, initially designed to meet specific communication needs among a large number of physicists, technicians and administrators who worked on high energy physics experiments, had a spectacular reach and is currently accessed by more than 4.5 billion people in the World.

In this scenario, however, there is a highly significant factor, which will mark a strong qualitative difference between the \textit{modern sociological subject} and the \textit{postmodern subject} of late 20th century and the beginning of the 21st century. It is the exaggerated \textit{fragmentation} of everything, it is what can be called the \textit{decentralization of the subject}. The perception that this new subject has of the World and of himself is built in an imaginary space of virtually infinite dimensionality, created by the internet and social networks.

It is the virtually unlimited www network that generates this imaginary, symbolic space, which, for all intents and purposes, has an infinite dimension. In it, Man gets lost. Pascal's perplexity in the face of an infinite (real) space is now repeated, in another context: the experience of facing an imaginary (abstract) space effectively infinite. The fractal character of the network also stands out. Space and time are fragmented. All Internet stories and narratives are incomplete instant flashes of the World, with no past or future. Man completely dissolves into what the Polish sociologist Zygmunt Bauman called \textit{liquid modernity}.

In our opinion, the emergence of \textit{postmodern subject}, using Stuart Hall's expression, or \textit{liquid subject}, in Bauman's nomenclature, is due to this almost tangible liquidity and the impact of \textit{infinity}, with a difference: while the \textit{enlightenment subject} was forged by facing the infinitude of the Cosmos and the real three-dimensional space, the postmodern subject, our contemporary, is characterized by the ease with which he gets lost in an infinitude of spaces in the network, with the speed of pressing an \textit{enter} key. Or, if we prefer, virtual multidimensional spaces come to have reality in the practical life of the individual. We are referring here to the basic characteristic of several concatenated hypertexts that allow that, at each \textit{web page}, the navigator can open another and another page and so on, practically without limits. Such text fragmentation, corresponding to a fragmentation of time and space, comes at a price. It is not hard to imagine that it can contribute to the dissolution of identity.

Indeed, it can be argued that the perception of \textit{cyberspace} ``not only develops a multiplicity of points of view, but also a set of selves''. Or that ``the individual self is giving way to the vague edges of identity'' \cite{Ascott}. According to the British artist Roy Ascott, there is a glimpse of a recurring redefinition, capable of creating multiple identities that operate in different places in cyberspace. This self, at the same time, multiplied, divided and dispersed, seems to be fundamental to life in the \textit{net}. It seems quite evident to us that these essential characteristics of this new subject force him to pulverize his own history. This kind of permanent denial of his roots and his history can lead to a corrosion of his own character, to which we are all exposed today. This is particularly true if we take into account the new labor relations in savage capitalism such as that seen in modern economy. We refer here to the point of view of the American sociologist and historian Richard Sennett~\cite{Sennett}. Let us see, albeit very briefly, how this process can take place.

For Sennett,
\begin{quotation}
\baselineskip=11pt{{\small \noindent ``character development depends on stable virtues such as loyalty, trust, commitment and mutual help. Features that are disappearing in the new capitalism. In some
respects, the changes that mark this new system are positive and have led to a dynamic economy, but they have also eroded the idea of purpose, integrity and trust in other aspects, aspects that previous generations considered essential for character formation.''}}
\end{quotation}

Its initial premise is that the motto of survival in the modern economy can put people's emotional lives adrift. We are all exposed to a strong tendency to ``decisively and irrevocably reinvent institutions, so that the present becomes discontinuous with the past''. Thus, it is justified the difficulty for individuals to build their own stories, based on their professional experiences and their ties of dependence with other individuals. Not to mention the decentralization of the subject which we have just referred to. A direct consequence of this scenario is that people tend to live only in the present.

Dreaming becomes more difficult, when the uncertainties of maintaining what has been achieved professionally become significant and when there is no ``deep mirror'' before us. The enormous flexibility of work, on the one hand, seeks to adapt quickly to the growing social changes and the immense volatility of consumer demand and, on the other hand, implies the acceptance that ``there is no long term''. Both are characteristics of a society that values consumption more than citizenship. We believe, like Sennett, that this expression contains the principle of corrosion of values such as trust, loyalty and mutual commitment. In this way, the spectrum of fragmentation and volatility is widening, extending beyond the borders of Economics, and infiltrating family, social relations and, especially, the School.

This postmodern subject, who emerged in the second half of the 20th century, but who irreversibly affirms him/herself, as we stress, from the generalization of the internet and the emergence of social networks, is contemporary with radical changes in Science and Culture. We are referring to the strong presence in Physics and Chemistry -- with Prigogine, for example -- of the study of chaos theory and complex systems, seen no longer as reducible to a sum of simple systems, in the way imagined by Descartes. This fact has contributed to a new attitude towards the world. We do not know to where these new paths of Science will lead, by valuing indeterminacy, non-linearity and complexity. However, there is no doubt that this trend is contemporary and, in a sense, can be seen as a reflection of postmodern subject. Perhaps a broader understanding of this complex subject only results from a new psychoanalysis, built from these new scientific paradigms, as has already been suggested~\cite{Palombo}.

Finally, we must emphasize that the non-existence of an unified Physics (like Aristotle's and Newton's) can be not only an open problem in Science but a problem in which a solution would be a possible requirement for a future Enlightenment, to believe in the lessons of the past. We look forward to the day when Cosmology and Particle Physics -- the old question of describing Physics in Heaven and Earth -- will be unified in a new theory of the Cosmos, an achievement that would weaken the current wave of irrationalism and pseudoscience. In any case, it is difficult to imagine the impact of this new Enlightenment on man.

\section{South American indian Culture: Plains and Cordil\-leras}\label{indios}

\begin{flushright}
\begin{minipage}{5.7cm}
\baselineskip=10pt
\vspace*{0.2cm}

\small{\textit{How now, Horatio? You tremble and look pale. Is not this something more than phantasy?} \\
\hspace*{2.7cm} \,Shakespeare~\cite{Shakespeare}}
\vspace*{0.15cm}
\end{minipage}
\end{flushright}

In this brief Section, we summarize some ideas that we hope to be able to develop elsewhere. In particular, we intend to argue that many aspects of Amerindian culture can be clarified if we try to see them under the light of their space ideas, especially those concepts connected to the experience of space dimensionality, that is, space verticality, given the geographical, environmental and ecological aspects of their habitats.

Since the 19th century, under the influence of evolutionary ideas, cultural anthropologists used to refer to native and indigenous peoples of  Africa, Oceania and the Americas as primitives or aborigines. In spite of isolated early criticism, this situation only began really to change after the seminal work by Levi-Strauss \cite{Levi-1,Levi-2,Levi-3}, in the mid 20th century, and seems to be now practically outmoded, at least in the academic world. The study of Brazilian indian Culture has played a fundamental role in these changes.

More recently, advances in Cultural Anthropology have greatly expanded the practical and empirical knowledge of Brazilian indian cultural life. See, for instance, Refs.~\cite{Vilaca,Fausto,Lima}. New and important theoretical points of view have also appeared after 1968 events~\cite{Clastres-1,Clastres-2,Clastres-3}.

Perhaps the most interesting and central point of Clastres' view is his idea that there are societies not only without State but, in fact, organized against the possibility of State. Clastres studies the Guayakis, an indian people living without State of their own in the plains of Paraguay, not far from Brazil. As a matter of fact, the Guayakis recognize chiefs in their settlements. However, these chiefs \cite{pouvoir} have no coercive power, limiting themselves to counselling, by exerting spiritual power in a conciliatory way, basically to avoid coercion and to prevent the emergence of coercive structures, especially those of organized State. Chiefs are chosen in complicated rituals and must exhibit certain leadership qualities, besides adequate kinships. In Xingu societies, similar in many aspects to the Guayakis, pubertal reclusion and excellence in \textit{huka-huka} fights are considered to be very important points. As a consequence, there are no social hierarchies, no privileged classes accumulating wealth. Furthermore, these societies are characterized by the absence of concentration of power mechanisms. This means that separate settlements remain territorially and demographically restricted, resisting unification. Clastres' remarkable conclusion is that the Guayakis are not simply living without State or even in a pre-State formation: they are, in fact, actively engaged against State. War in this society does not aim for the expansion of Power or for the formation of an Empire. On the contrary, war is waged to guarantee the real Independence of the settlements, to prevent the formation and existence of organized State, \textit{i.e.}, coercive Power.

Now, what calls our attention is the great contrast between the remarkable situation of these people, living in the plains, and the well known presence in the Continent of the highly organized Inca State in the mountains, the Andes Cordillera.

This situation suggests a simple interpretation: humans living in the high mountains of the Cordillera naturally develop a sense of verticality, associated to hierarchy. The great mountains, like monuments created by Nature, inspire the sense of awe and respect which we feel before a medieval cathedral. Here, it is important to stress how the cathedral constructions were used to reinforce the power of the Church and to forge the medieval subject and his relationship to a theocentric society.

This concrete and unforgettable experience of organized matter in space through its verticality is the driving force of hierarchies and their metaphors: up and down. Humans living in the Cordilleras, contemplating verticality, are prone to imagine, create, build and accept social hierarchies, organized State and Empire, which emerge as natural images in the mirror of an imposing environment. On the other hand, the Imaginary of the Guayakis flies freely in the direction of Independence, absence of  hierarchies and organized State, \textit{i.e.}, coercive power. This link between the living experience of Space and human imagination is a theme which Bachelard deeply explored in his \textit{Po\'{e}tique de l'espace}~\cite{Bach-espace}. Now, our suggestion should not be considered a deterministic point of view: the experience of environment and space only opens possibilities, that is all.

A similar situation can be observed in the lowlands of Amazon. European and civilized people have constructed a metaphysical description of the world in which Nature is an invariant, it is the same for all the peoples of the world, while Culture may vary. This is, of course, \textit{multiculturalism}. However, the point of view of the Amazonian indians is quite the opposite. For them, Nature is variable and Culture is invariant, being the same not only for all men and women, but in fact for all living animal and even spirits, which are essentially humans. From the viewpoint of the Amazonian indians, every species considers itself, and must be considered, to be human and sees the others including us as non-humans. Ethnographic evidence supports this view, for instance, the studies by Phillipe Descola~\cite{Descola}, among others.  This is called \textit{multinaturalism}, in sharp contrast with traditional multiculturalism, and, actually, according to Viveiros de Castro~\cite{Viveiros-1,Viveiros-2,Viveiros-3,Viveiros-4}, corresponds to a new metaphysical approach: the Amazonian \textit{perspectivism}. The logic underlying this view is based on the prey $\times$ predator relationship.
The essential point is that, for the Amazonian indian there is no Hierarchy in Nature. Humans and animals are ontologically the same. This may possibly indicate that, for Amazonian indians, verticality does not play a fundamental role in the formation of their Imaginary. Their \textit{Weltanschauung} is essentially two-dimensional like that of the Guayakis.

Another example may be extracted from the Bororo of Brazil, living in the state of Mato Grosso. ``Although their village, patterned by cosmological concepts, has an ``upper'' portion and a ``lower'' portion this does not imply a hierarchical system. Indeed, for the Bororo, ``the terms <<upper>> and <<lower>>, while descriptive of an ideal topographic/cosmological feature and applicable as such to the village social units, cannot be taken as associated in any simple fashion with relative degrees of social order''~\cite{Fabien}.

Our final remark is that it should be strongly emphasized that we are not preaching geographical determinism in History. We know that, in general, ancient human groups evolved towards hierarchical societies. This represents the usual situation around the world and has been discussed and clarified by many authors. Consider, for instance, the Egyptian case. They live in a land without nearby mountains and developed a highly hierarchical society. However, their myths and legends were dominated by the presence of the Nile. Usually, they never explored the hostile desert around. The Imaginary of this extraordinary civilization  was molded by the River and its cycles. Herodotus knew that. This seems to be the situation of most ancient civilizations in Mesopotamia (Tigris-Eufrates), China (Yellow and Yangtse), and possibly also India and Pakistan (Ganges and Indus) in spite of the imposing presence of the Himalaya. Our aim is very modest and simple: we are trying to understand what seems to be South American exception. We are only saying that Human Imaginary may be fed by the living conditions of the people, their natural and spiritual environment -- \textit{Physis} and \textit{Psyche} -- opening unexpected dialectic possibilities to History. This could possibly contribute to clarify unexpected situations like that of the Guaiakis or Ach\'{e}. Pierre Clastres would possibly say: L'imagination au pouvoir.

We are aware of the speculative and incomplete features of the ideas exposed in this Section. There are certainly many other important factors that should be taken into account to fully understand the relation between hierarchy and space concepts in a certain society, like its cosmogony, for example. However, knowing Doria well, we are convinced that he will understand and appreciate our daring to put them on paper. Whether these conjectures will be worked out and eventually bear fruit or not, only the future will tell us.

\section{Concluding remarks}\label{final}

The general question of the subject and his relations with Culture were dealt with here. To recap, we saw that Christianity's adoption of \textit{codex} is linked to the issue of \textit{verticality} or, in other words, to
\textit{space three-dimensionality}. It is through this third dimension that the \textit{medieval subject} is related to God. The invention of this object, the \textit{codex}, in the High Middle Ages occupies a prominent place in the formation of the emerging ``three-dimensional'' theocentric Man. The essence of the \textit{codex}, which is its three-dimensionality, will be associated with culture and its transmission for a long period.

We have also argued that a new Man emerges from the Renaissance, who came to be called by Stuart Hall \textit{enlightenment subject}, who, after all, results from an inextricable entanglement of concepts which are mirrored in the crucial question of the relationship of Man with the Cosmos. This implies two notable capacities of this subject: that of coming to conceive the \textit{infiniteness} of the Universe and of three-dimensional space and that of not succumbing to the amazement caused by this non-divine infinity. He does not become small. On the contrary, as has already been seen, he grows, from the moment when he realizes how much, through his intellect, he is perfectly capable of embracing, conceiving and rationalizing all these questions that touch infinity, from the microcosm to the macrocosm.

We found that the great narratives about Man, which ultimately define our view of ourselves -- that mirror that responds -- are intertwined with those concepts that Cassirer identified as the foundation of culture: \textit{space}, \textit{time}, and \textit{number}. The uni-dimensionality of time -- associated with the Christian Worldview -- is reinforced by the linearity of the text in the \textit{codex}, which does not require repetition.

Indeed, in this essay, we suggest that the dimensionality of the writing support -- \textit{volumen}, \textit{codex}, book -- may have a major influence on Culture: bidimensionality, in the case of \textit{volumen}, and the three-dimensionality of \textit{codex}. On the other hand, the postmodern subject, as previously discussed, is characterized by the multi-dimensionality of digital networks, which tends to infinity. \textit{Number}, on the other hand, manifests itself in the spatialization of Algebra -- with the emergence of Analytical Geometry -- and in the mathematization of the physical description of the World.

We are tempted to conclude that it is the dimensionality of the imaginary space in which Man lives, linked to the reading of texts and of the World, the main key to understand, in each epoch, the subject's striking features.
As a new application of these ideas, we sketched, in Section~\ref{indios}, the possibility of understanding some cultural and social differences between two different indian groups from their different space conceptions.

At this point, we cannot resist the temptation to quote how Albert Camus refers to the problem of the dimensionality of space in the opening of his book \textit{Le Mythe de Sisyphe}, in which he deals with the issue of the human being becoming aware of meaninglessness of his condition (absurd Man faces an absurd world and humanity):
\begin{quotation}
\baselineskip=11pt{{\small \noindent ``There is only one really serious philosophical problem: it is suicide. To judge whether life is worth living or not is to answer the fundamental question of philosophy. The rest, if the world has three dimensions, if the spirit has nine or twelve categories, appears next.''}}
\end{quotation}

We would now like to end the paper with an open question.

If it is true that the central issue of \textit{subjectivity} in Post-Modernity (or Liquid Modernity) can be understood as related to the infinitude of dimensions of new virtual spaces, is it possible to establish a certain analogy with what happened in the Renaissance, when Man lived astonished by the infinitude of space and Cosmos? This leads us to an inescapable question, which we leave to the reader. Will there be a new Enlightenment in the foreseeable future?

If we had to risk an answer, we would first remember that 18th century Enlightenment was only possible after the overcoming of the crisis introduced by Copernicus in Science, by breaking the unity of Aristotelian Cosmos. Newton did it from the moment he theoretically and conceptually reunited Physics and Astronomy. That said, and reasoning by analogy, we believe that it is practically unlikely that, without a qualified theory unifying General Theory of Relativity and Quantum Theory of Matter, the credibility of Science, and the belief in Reason, which is indispensable to the new Enlightenment, will be restored.

\section*{Acknowledgment}

We would like to express our thanks to Jos\'{e} Acacio de Barros for the kind invitation to contribute to Francisco D\'{o}ria's \textit{Festschrift}, which is an honor for us. We are in debt to H\'{e}lio da Motta Filho, Felipe Silveira, and Pedro Teixeira who carefully read the manuscript, and to M\'{e}rcio Pereira Gomes and Adelino de Lucena Mendes da Rocha for fruitful discussions on a few anthropology matters and for their critical remarks, although we dared to disagree in some important issues. However, some points raised by them will deserve our attention in the near future. In any case, needless to say that they are not responsible for our speculative approach.

\renewcommand\refname{Notes and References}


\begin{thebibliography}{50}

\bibitem{Cassirer-Man}
Ernst Cassirer. \textit{Philosophie des formes symboliques. 2: La pens\'{e}e mythique}. Paris: \'{E}dition Minuit, 1972.

\bibitem{nosso-livro}
Francisco Caruso \& Roberto Moreira Xavier. \textit{O Livro, o Espa\c{c}o e a Natureza: Ensaio Sobre a Leitura do Mundo, as Muta\c{c}\~{o}es da Cultura e do Sujeito}. S\~{a}o Paulo: Livraria da F\'{\i}sica, 2017.

\bibitem{Popper}
Karl Popper. \textit{Conjectures and Refutations:~The Growth of Scientific Knowledge}. Chapter~1, Section~VII. London: Routledge and Kegan Paul PLC, 1969.

\bibitem{Arkan}
Arkan Simaan \& Jo\"{e}lle Fontaine. \textit{L'image du monde. Des Babyloniens \`{a} Newton}. ADAPT Editions, 1999.

\bibitem{Apud}
\textit{Apud} Ernst Cassirer. \textit{An Essay on Man}. New Haven, Connecticut: Yale University Press, 1944.

\bibitem{Bachelard}
Gaston Bachelard. \textit{La terre et les r\`{e}veries du repos}. Paris: Librairie Jos\'{e} Corti, 1948.

\bibitem{Reale}
Giovanni Reale. \textit{Storia della filosofia greca e romana}, vol~IX. Milano: Bompiani, 2004.

\bibitem{Ladner}
Gerhardt B.~Ladner. \textit{God, Cosmos, and Humankind: The World of Early Christian Symbolism}. English translation by Thomas Dunlap. Berkeley, Los Angeles \& London: University of California Press, 1995.

\bibitem{Dijk}
 Eduard Jan Dijksterhuis. \textit{The Mechanization of the World Picture: Pythagoras to Newton}. Princeton: Princeton University Press, 1986.

\bibitem{Renasc}
Wallace K.~Ferguson. \textit{The Renaissance in Historical Thought: Five Centuries of Interpretation}. Toronto: University of Toronto Press, 2006.

\bibitem{Hall}
Stuart Hall. ``The question of cultural identity'', \textit{in} Stuart Hall, David Held \& Anthony McGrew (Eds.). \textit{Modernity and Its Futures}. Cambridge:~Polity Press, 1992,  p.~274-316.

\bibitem{Vico}
\textit{Cf.}, for example, Giambattista Vico. \textit{La scienza nuova}. Napoli: Muziana Stamperia, 1744.

\bibitem{Harvey}
An analysis of the meaning of postmodern conception in different contexts, which also seeks to identify its degree of accuracy and usefulness in describing contemporary experience, can be found in David Harvey.
\textit{Condition of Postmodernity: An Enquiry into the Origins of Cultural Change}. Cambridge, Massachusetts: Blackwell, 1992.

\bibitem{Fromm}
Erich Fromm. \textit{Marx's Concept of Man}. Continuum International Publishing Group Ltd., 1981.

\bibitem{Marx}
Karl Marx. \textit{Economic and Philosophic Manuscripts of 1844}. International Publishers Co, 1980.

\bibitem{Havelock}
Eric A.~Havelock. \textit{The Literate Revolution in Greece and its Cultural Consequences}. Princeton: University Press, 1982.

\bibitem{Havelock-Plato}
Eric A.~Havelock. \textit{Preface to Plato}. Cambridge, MA: Harvard University Press, 1963, Chapter~X.

\bibitem{Havelock-Muse}
Eric A.~Havelock.~\textit{The Muse Learns to Write.~Reflections on Orality and Literacy from Antiquity to the Present}. New Haven and London: Yale University Press, 1986.

\bibitem{Jacob}
Christian Jacob. ``Lire pour \'{e}crire: navigations alexandrines'', in Marc Baratin \& Christian Jacob (Eds.). \textit{Le Pouvoir des Biblioth\`{e}ques: La M\'{e}moire des Livres en Occident}. Paris: Albin Michel, 1996, p.~53.

\bibitem{Roberts}
Colin H.~Roberts \& T.C.~Skeat. \textit{The Birth of the Codex}. London: Oxford University Press, 1983.

\bibitem{Cavallo}
Guglielmo Cavallo. ``Testo, libro, lettura'', \textit{in} Gugliemo Cavallo, Paolo Fedeli \& Andrea Giardina (Eds.). \textit{Lo spazio letterario di Roma antica}, v.~II, \textit{La circolazione del testo}. Roma: Salerno Editrice, 1989, pp.~307-341; ``Libro e cultura scritta'', \textit{in} \textit{Storia di Roma}, v.~4, \textit{Caratteri e Morfologie}. Torino: Einaudi, 1989, pp.~693-734.

\bibitem{Bruc}
From Jacques Le Goff's Preface to the book of Philippe Bruc. \textit{L'Ambigu\"{\i}t\'{e} du livre: Prince, pouvoir, et peuples dans les commentaires de la Bible au Moyen Age}. Paris: Beauchesne \'{E}diteur, 1994.

\bibitem{Ursula}
\'{U}rsula E.~Katzenstein. \textit{A origem do Livro: da Idade da Pedra ao Advento da Impress\~{a}o Tipogr\'{a}fica no Ocidente}. S\~{a}o Paulo: Editora HUCITEC e Instituto Na\-cio\-nal do Livro, 1986, p.~37.

\bibitem{McLuhan}
Marshall McLuhan. \textit{Understanding Media: The Extensions of Man}. McGraw-Hill Paperbacks, 1965.

\bibitem{Cassirer-Vesalius}
Ernst Cassirer. ``The place of Vesalius in the Culture of Renaissance''. \textit{Yale Journal of Biology and Medicine}, v.~16, n.~2, pp.~109-120 (1943).

\bibitem{Huizinga}
On the hierarchical conception of medieval society see Chapter~3 of the classic book of Johan Huizinga. \textit{The Waning of the Middle Ages: A Study of the Forms of Life, Thought and Art in France and the Netherlands in the XIVth and XVth Centuries}. New York: St. Martin's Press, 1984.

\bibitem{Lilley}
Keith D.~Lilley. \textit{City and Cosmos: The Medieval World in Urban Form}. UK:~Reak\-tion Books, 2009.

\bibitem{Pano}
Regarding the analogies between art, philosophy and theology in the architecture of the Middle Ages, we refer the reader to the classic of the critic and historian of German art Erwin Panofsky. \textit{Gothic Architecture and Scholasticism}. New York: The World Publishing Company, 1968.

\bibitem{LeGoff}
Jacques Le Goff.~\textit{La Naissance du Purgatoire}.~Paris: \'{E}di\-tions Gallimard, 1981.

\bibitem{Koyre}
Alexandre Koyr\'{e}. \textit{La R\'{e}volution Astronomique: Copernic, Kepler, Borelli}. Paris: Hermann Impr.~Union, 1961.

\bibitem{Rossi}
Paolo Rossi (Ed.). \textit{La Rivoluzione Scientifica: da Copernico a Newton}. Torino: Loescher Editore, 1973, p.~125.

\bibitem{Duhem}
For a history of the theories of the infinite and the plurality of worlds see: Pierre Duhem. \textit{Medieval Cosmology: Theories of Infinity, Place, Void, and the Plurality of Worlds}. Chicago and London: The University of Chicago Press, 1985.

\bibitem{Pascal}
Blaise Pascal. \textit{Pens\'{e}es}. Paris: Guillaume Desprez, 1670.

\bibitem{ordre}
Roger Chartier. \textit{L'Ordre des Libres: Lecteurs, Auteurs, Biblioth\`{e}ques en Europe entre XIVe et XVIIIe si\`{e}cle}. Aix-en-Provence: Alin\'{e}a, 1992.

\bibitem{redes}
An extensive analysis of the social and economic impact of the networks was done by Manuel Castells. \textit{The Rise of the Network Society}. Malden, MA, and Oxford: Blackwell Publishers, 1996.

\bibitem{Bloch}
R.~Howard Bloch \& Carla Hesse.~\textit{Future Libraries}. Berkeley and Los Angeles:~University of California Press, 1995.

\bibitem{Lister}
Martin Lister (Ed.). \textit{The Photographic Image in Digital Culture}. London: Routledge, 1995.

\bibitem{Ascott}
Roy Ascott. ``The Architecture of cyberception''. \textit{In} M.~Toy (Ed.). \textit{Architectural design. Architects in cyberspace}, v.~65, n.~11/12. London: Academy Editions, 1995, pp.~38-41.

\bibitem{Sennett}
Richard Sennett. \textit{The Corrosion of Character: The Personal Consequences of Work in the New Capitalism}. New York: W.W. Norton \& Company, 1998.

\bibitem{Palombo}
Stanley R.~Palombo. \textit{The Emergent Ego: Complexity and Coevolution in the Psychoanalytic Process}. Madison, Connecticut: International Universities Press, 1999.

\bibitem{Shakespeare}
William Shakespeare. \textit{Hamlet}, act~1, scene 1.

\bibitem{Levi-1}
Claude Levi-Strauss. \textit{Tristes tropiques}. Paris: Librairie Plon, 1955.

\bibitem{Levi-2}
Claude Levi-Strauss. \textit{Anthropologie Structurale}. Paris: Librairie Plon, 1958.

\bibitem{Levi-3}
Claude Levi-Strauss. \textit{Pens\'{e}e Sauvage}. Paris: Librairie Plon, 1958.

\bibitem{Vilaca}
Aparecida Vila\c{c}a. \textit{Palet\'{o} e eu: mem\'{o}rias de meu pai ind\'{\i}gena}. S\~{a}o Paulo: Editora Todavia, 2018.

\bibitem{Fausto}
Carlos Fausto \& L.A.~Lino da Costa. ``Recent Studies of Amazonian Ontologies''. \textit{Religion and Society}, v.~1, n.~1, pp.~89-109, 2011.

\bibitem{Lima}
T\^{a}nia Stolze Lima. \textit{Um peixe olhou para mim: o povo Yudj\'{a} e a perspectiva}. S\~{a}o Paulo: Editora UNESP, 2011.

\bibitem{Clastres-1}
Pierre Clastres. \textit{Chronicle of the Guayaki indians}. New York: Zone Press, 1998.

\bibitem{Clastres-2}
Pierre Clastres. \textit{La societ\'{e} contre l'\'{E}tat: recherches d'Anthropologie Politique}. Paris, \'{E}ditions de Minuit, 1974.

\bibitem{Clastres-3}
Pierre Clastres. \textit{Archeology of Violence}. New York: Semiotext(e), 1994.

\bibitem{pouvoir}
Pierre Clastres. \'{E}change et pouvoir: philosophie de la chefferie indienne. \textit{L'Hom\-me}, v.~2, n.~1, pp.~51-65, 1962.

\bibitem{Bach-espace}
Gaston Bachelard. \textit{Po\'{e}tique de l'espace}. Paris: PUF, 1957.

\bibitem{Descola}
Philippe Descola. \textit{Beyond Nature and Culture}. Chicago: University of Chicago Press, 2005.

\bibitem{Viveiros-1}
Eduardo Viveiros de Castro. ``O m\'{a}rmore e a murta: sobre a inconst\^{a}ncia da alma selvagem''. \textit{Revista de Antropologia}, v.~35 (dezembro) p.~21, 1992.

\bibitem{Viveiros-2}
Eduardo Viveiros de Castro. \textit{A inconst\^{a}ncia da alma selvagem}. S\~{a}o Paulo: Cosac \& Naify, 2002.

\bibitem{Viveiros-3}
Eduardo Viveiros de Castro. \textit{Metaf\'{\i}sicas canibais: elementos para uma antropologia p\'{o}s-estrutural}. S\~{a}o Paulo: Ubu Editora, 2009.

\bibitem{Viveiros-4}
Eduardo Viveiros de Castro. \textit{Perspectivismo e multinaturalismo na Am\'{e}rica Ind\'{\i}gena}. S\~{a}o Paulo: Ubu Editora, 2018.

\bibitem{Fabien}
Stephen M.~Fabien. \textit{Space-time of the Bororo of Brazil}. Gainesville: University of Florida, 1992.
\end{thebibliography}
\end{document}